\documentclass[10pt,twocolumn,oneside,letterpaper,conference]{IEEEtran}
\usepackage{cite}
\usepackage{subfigure}
\usepackage{graphicx}
\usepackage{epstopdf}
\usepackage{amsmath}
\usepackage{txfonts}
\usepackage{times}
\usepackage{latexsym}
\usepackage{bm}
\usepackage{amssymb}
\usepackage[center]{caption2}
\usepackage{stfloats}
\usepackage{cases}
\usepackage{array}
\usepackage{setspace}
\usepackage{fancyhdr}
\usepackage{balance} 
\usepackage{multirow}
\usepackage{booktabs}
\usepackage{amsmath}
\usepackage{algorithm}
\usepackage{algpseudocode}
\usepackage{CJK}
\usepackage{xcolor}
\usepackage{amsmath}
\usepackage{booktabs}
\usepackage{multirow}
\usepackage{array}
\usepackage{bbm}

\allowdisplaybreaks[4]

\newcaptionstyle{mystyle1}{%
	\centering TABLE  \captiontext \par}
\captionstyle{mystyle1}
\newcaptionstyle{mystyle2}{%
	\captionlabel $.$ \: \captiontext \par}
\captionstyle{mystyle2}
\newcaptionstyle{mystyle3}{%
	\captionlabel $.$ \: \captiontext \par}
\captionstyle{mystyle3}

\begin{document}

\title{\huge MDS Codes Based Group Coded Caching in Fog Radio Access Networks}

\author{
	\IEEEauthorblockN{Qianli Tan$^{1}$,
		Yanxiang Jiang$^{1,2,*}$,
		Fu-Chun Zheng$^{1,2}$,
		Mehdi Bennis$^{3}$,
		and Xiaohu You$^1$}
	\IEEEauthorblockA{$^1$National Mobile Communications Research Laboratory,
		Southeast University, Nanjing 210096, China.\\
		$^2$School of Electronic and Information Engineering, Harbin Institute of Technology, Shenzhen 518055, China.\\
		$^3$Centre for Wireless Communications, University of
		Oulu, Oulu 90014, Finland.\\
		E-mail: $\{$ 220200706@seu.edu.cn, yxjiang@seu.edu.cn, fzheng@ieee.org, mehdi.bennis@oulu.fi, xhyu@seu.edu.cn $\}$
}}

\maketitle

\begin{abstract}
In this paper, we investigate maximum distance separable (MDS) codes based group coded caching in fog radio access networks (F-RANs). 
The goal is to minimize the average fronthaul rate under nonuniform file popularity. 
Firstly, an MDS codes and file grouping based coded placement scheme is proposed to provide coded packets and allocate more cache to the most popular files simultaneously. 
Next, a fog access point (F-AP) grouping based coded delivery scheme is proposed to meet the requests for files from different groups.
Furthermore, a closed-form expression of the average fronthaul rate is derived. 
Finally, the parameters related to the proposed coded caching scheme are optimized to fully utilize the gains brought by MDS codes and file grouping. 
Simulation results show that our proposed scheme obtains significant performance improvement over several existing caching schemes in terms of fronthaul rate reduction.	
\end{abstract}

\begin{IEEEkeywords}
Coded caching, fog radio access networks, nonuniform popularity distribution, maximum-distance separable codes, fronthaul rate
\end{IEEEkeywords}

\section{Introduction}
The rapid development of mobile technology and increasing number of mobile devices have led to unprecedented growth in network traffic \cite{wang2017integration}. 
Fog radio access network (F-RAN) has emerged as a promising architecture to deal with this challenge \cite{peng2016fog}. 
In F-RANs, fog access points (F-APs) are connected to the cloud server through fronthaul links and equipped with caching resources. 
F-APs are close to users and able to provide higher quality of service (QoS) by using their caching resources. 
In addition, caching has emerged as an effective technique to reduce peak traffic by pre-storing parts of the popular contents in users' local caches during off-peak traffic time.

In \cite{maddah2014fundamental}, a new caching scheme called coded caching was proposed to further reduce the network congestion by elaborately designing the content placement phase and the content delivery phase. 
In \cite{6807823}, a decentralized setting of coded caching was introduced, where users  cache a certain number of bits of files randomly and independently.
Additional research on coded caching has extended to topics such as asynchronous coded caching \cite{8632748}, hierarchical coded caching \cite{karamchandani2016hierarchical}, online coded caching \cite{pedarsani2015online} and private coded caching \cite{ravindrakumar2017private}.

The use of maximum distance separable (MDS) codes in coded caching was studied in \cite{8525434, 9145207, wei2017novel, reisizadeh2018erasure}.
MDS codes can provide redundant packets to reduce the transmission load in caching networks.
The authors in \cite{8525434} used the scheme in \cite{6807823} to construct coded multicasting messages and further encode the multicasting messages by MDS codes. 
In \cite{9145207}, the authors investigated the employment of MDS codes in constructing coded multicasting messages for requests on the same file and different files. 
In \cite{wei2017novel} and \cite{reisizadeh2018erasure}, an MDS codes based decentralized coded caching was proposed, where the server first encodes each file by MDS codes and then performs random placement and delivery on the MDS coded files. 

The previous works in \cite{8525434, 9145207, wei2017novel, reisizadeh2018erasure} all assume the file popularity to be uniformly distributed. 
Nevertheless, the file popularity distribution is often nonuniform in real scenarios. 
The performance of coded caching under nonuniform file popularity was studied in \cite{niesen2016coded, ji2017order, zhang2017coded}. 
In \cite{niesen2016coded}, the authors divided the files in the database into multiple groups, where files in the same group have similar popularity and each group is assigned a separate portion of the cache. 
In \cite{ji2017order}, a decentralized coded caching scheme was proposed, where files are divided into only two groups, and the gap between the achievable bound and the lower bound is shown to be a constant only when the file popularity follows a Zipf distribution. 
In \cite{zhang2017coded}, the authors proposed a coded caching scheme similar to the scheme in \cite{ji2017order}, where files are also divided into two groups, and the achievable average load is within a constant factor of the minimum average load under an arbitrary file popularity distribution. 
However, the works in \cite{niesen2016coded, ji2017order, zhang2017coded} mainly focus on designing the coded caching scheme with an uncoded placement phase and a coded delivery phase, where files are not encoded by erasure codes such as MDS codes in the placement phase. As a result, the benefits of coded placement have not been well studied yet.

Motivated by the aforementioned discussions, it is important to study the benefits of coded placement when the file popularity is nonuniformly distributed. 
To reduce the average fronthaul rate in F-RANs, we propose an MDS codes based group coded caching scheme.   
Based on the closed-form expression of the fronthaul rate, we optimize the parameters related to the proposed scheme.
In our proposed scheme, the cloud server can benefit from the reconstruction property of MDS codes and the gain brought by allocating more cache for the most popular files simultaneously. 
As a result, the average fronthaul rate can be significantly reduced. 

The rest of this paper is organized as follows. 
In Section II, the system model is described. 
In Section III, we present our proposed MDS codes based group coded caching scheme. 
Simulation results are shown in Section IV. 
Finally, conclusions are drawn in Section V.

\section{System Model}
Consider the F-RANs depicted in Fig. 1, where the cloud server is connected to $K$ F-APs through a shared and error-free fronthaul link.  
The cloud server has access to a database of $N$ files, namely $W_1,W_2,\ldots,W_N$. 
The size of each file is $F$ bits.
Let ${\cal K}=\{1,2,\ldots,k,\ldots,K\}$ denote the index set of $K$ F-APs, and ${\cal N}={\{1,2,\ldots,n,\ldots,N\}}$ the index set of $N$ files.
Each F-AP is equipped with a cache whose capacity is $MF$ bits.
Assume that each F-AP only serves a single user at a time.
For each F-AP, when its served user requests a file, it will inform the cloud server of the request immediately.
For description convenience, we refer to the above process as an F-AP sending a request for a file. 
Let $\boldsymbol{p} = \left[p_1,p_2,\ldots,p_N \right]^T$ denote the file popularity distribution, where $\sum_{n=1}^{N}p_n = 1$. 
Without loss of generality, these files can be relabeled according to their popularity and we assume $p_1 \ge p_2 \ge \ldots \ge p_N$.
The popularity of $N$ files follows a Zipf distribution, and the popularity of file $W_j$ is defined as follows:
\begin{equation}
	p_j = \frac{j^{-\alpha}}{\sum_{n=1}^{N} n^{-\alpha}},
\end{equation}
where $\alpha$ is the parameter of the Zipf distribution, and its typical value is between 0.5 and 2, i.e., $\alpha \in \left[ 0.5, 2 \right]$. 

The system operates in two phases. 
In the placement phase, the caches of the F-APs are filled with contents from the database. 
In the delivery phase, each F-AP sends its request for a file independently.
Let $\boldsymbol{d} = \left[d_1,d_2,\ldots,d_K \right]^T$ denote the request vector of the users, where $d_k$ is the index of the file requested by F-AP $k$. 
Based on the received requests and the cached contents of F-APs, the cloud server forms coded multicasting contents and then transmits them to all the F-APs.

Let $\varphi$ denote the placement strategy.  
Let $R_{\varphi,\boldsymbol{d}}$ denote the fronthaul rate corresponding to placement strategy $\varphi$ and request vector $\boldsymbol{d}$.
Suppose that the cloud server needs to transmit $R_{\varphi,\boldsymbol{d}}F$ bits of coded multicasting contents to the F-APs to guarantee that each F-AP can reconstruct its requested file under placement strategy $\varphi$ and request vector $\boldsymbol{d}$.
In order to analyze the performance of our proposed scheme from a practical perspective, we focus on the average fronthaul rate. 
It is defined over all possible request vectors as follows:

\begin{figure}[!t]
	\centering
	\includegraphics[width=0.45\textwidth]{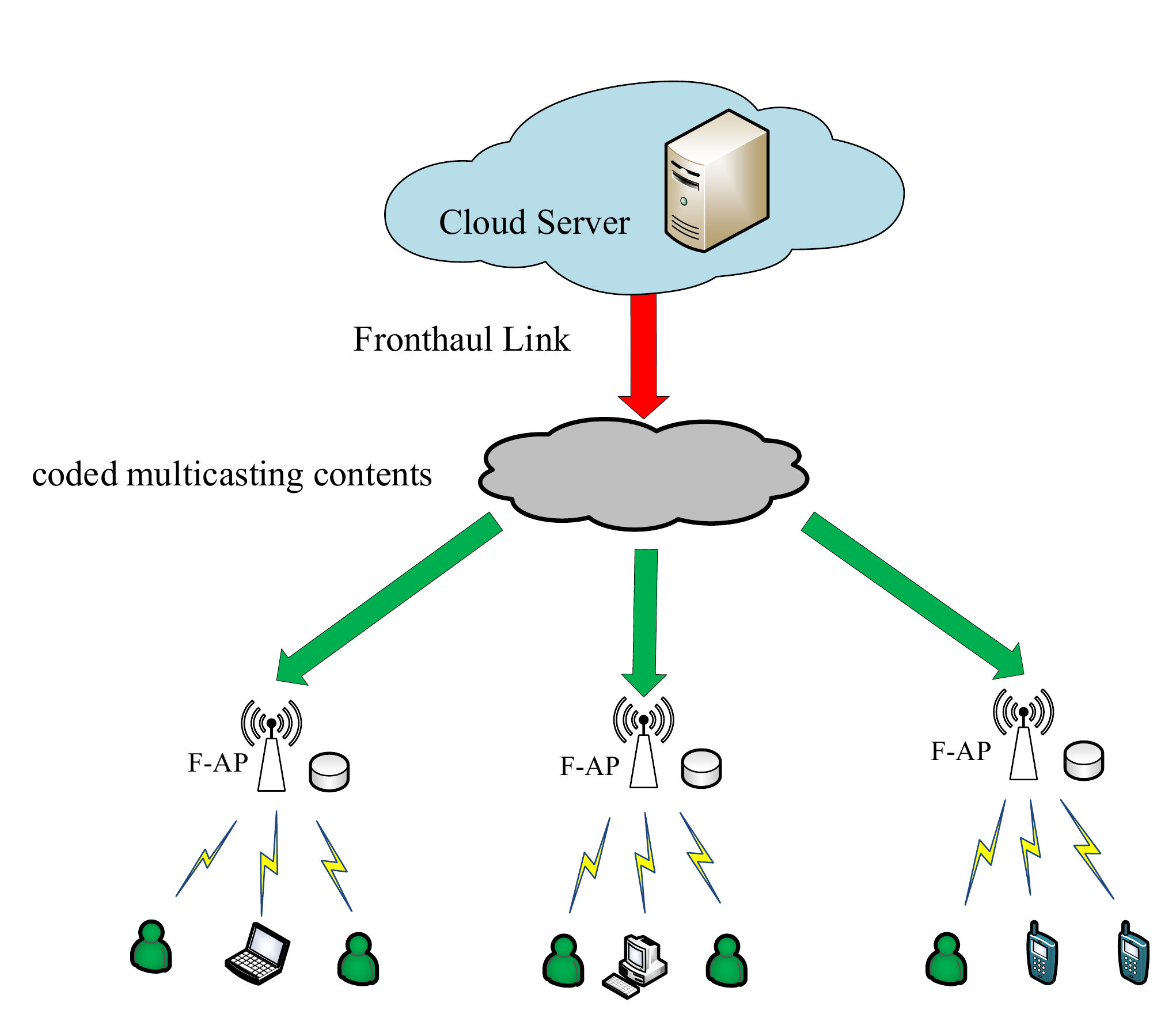}
	\caption{Illustration of the coded caching scenario in the F-RANs.}
	\label{fig1}
\end{figure}

\noindent \textbf{Definition 1.} The average fronthaul rate of a coded caching scheme under placement strategy $\varphi$ is defined as 
\begin{equation}
	\overline{R}_{\varphi}(K,M,N) = \sum_{\boldsymbol{d} \in \Theta} \left( \prod_{k=1}^K p_{d_k} \right) R_{\varphi,\boldsymbol{d}}(K,M,N),	
\end{equation}
where $\Theta$ is the set of all possible demand vectors, $|\Theta| = N^K$.

In this paper, we wish to minimize the average fronthaul rate by optimizing the placement strategy, and the problem is formulated as follows:
\begin{equation}
	\min_{\varphi} \overline{R}_{\varphi}(K,M,N).
	\label{optimization1}
\end{equation}
 
\section{Proposed MDS Codes Based Group Coded Caching Scheme}

In this section, we first propose the MDS codes and file grouping based coded placement scheme. 
Then, we propose the F-AP grouping based coded delivery scheme. 
Finally, we derive the closed-form expression of the average fronthaul rate and further optimize the parameters related to the placement scheme.

\subsection{Proposed MDS Codes and File Grouping Based Coded Placement Scheme} 
 
First, each file in the database is encoded by an $(F/r,F)$ MDS code, where $r$ is the rate of the MDS code with $r \in (0,1]$.
It is assumed that an MDS code of rate $r$ can map a file $W_i$ of $F$ bits into a coded file $W^{'}_i$ of $F/r$ bits, and any distinct $F$ bits out of the $F/r$ bits of $W^{'}_i$ is sufficient to reconstruct the original file $W_i$. 
Next, a file split point is selected. 
Let $N_0$ denote the file split point with $N_0 \in (0,N]$.
Based on the file split point, we propose to divide the coded files in the database into two groups, namely \emph{cached group} and \emph{uncached group}. 
Let ${\cal N}_{1}={\{1,2,\ldots,N_0\}}$ denote the index set of the files in the cached group, and ${\cal N}_{2}={\{N_0+1,N_0+2,\ldots,N\}}$ the index set of the files in the uncached group.
Then, we apply different caching strategies to these two file groups.
For the files in the cached group, each F-AP independently and randomly caches $MF/N_0$ bits of each file.
The files in the uncached group are not cached by any F-AP.

In brief, the placement scheme consists of three steps.
In the first step, each file is encoded by an MDS code.
In the second step, the MDS coded files are divided into two groups.
In the last step, each F-AP cache a certain number of bits of the files in the first group. 
The proposed coded placement scheme is presented in Algorithm 1.

\begin{algorithm}[!t]
\caption{MDS Codes and File Grouping Based Coded Placement Scheme}
\label{placement}
\begin{algorithmic}[1]
\For{$i \in {\cal N}$}	
\State Use an MDS code of rate $r$ to encode file $W_i$ into coded file $W_{i}^{'}$;
\EndFor
\State Split the file index set $\cal{N}$ into two subsets based on $N_0$: ${\cal N}_{1} = {\{1,2,\cdots,N_0\}}$ and ${\cal N}_{2} = {\cal N} \backslash {\cal N}_{1}$; 
\For{$k \in {\cal K}, i \in {\cal N}_1$}
\State F-AP $k$ independently and randomly caches $\frac{MF}{N_0}$ bits of coded file $W_{i}^{'}$;
\EndFor

\end{algorithmic}	
\end{algorithm}

\addtolength{\topmargin}{0.01in}
\subsection{Proposed F-AP Grouping Based Coded Delivery Scheme}

First, the F-APs are divided into two groups based on their file requests.
Let ${\cal K}_{1}$ denote the index set of F-APs that request a file in the cached group, and ${\cal K}_{2}$ the index set of F-APs that request a file in the uncached group. 
Note that F-AP $k$ can reconstruct its requested file $W_{d_k}$ after receiving $F$ bits of the coded file $W^{'}_{d_k}$. 
Therefore, for the F-APs in ${\cal K}_1$, since they have already cached $MF/N_0$ bits of their requested file, the cloud server keeps transmitting coded multicasting messages to them until all of them receive $1-MF/N_0$ bits of their requested file. 
For a subset of F-APs ${\cal P} \subseteq {\cal K}_1$, let $W^{'}_{d_k,{\cal P} \backslash \{k\}}$ denote the set of bits of $W^{'}_{d_k}$ that are exclusively cached by F-APs in ${\cal P} \backslash \{k\}$, $k \in {\cal P}$. 
The coded multicasting message $\oplus_{k \in {\cal P}} W^{'}_{d_k, {\cal P} \backslash \{ k\}}$ transmitted by the cloud server is simultaneously useful for $p$ users, where $p=|{\cal P}|$.
In particular, the cloud server starts with transmitting the coded messages that are useful for the largest number of F-APs, and then transmits the coded messages that are useful for fewer F-APs.
Due to the reconstruction property of MDS codes, the number of transmissions that are useful only for a small subset of ${\cal K}_1$ can be reduced.  
For the F-APs in ${\cal K}_2$, since their requested files are not cached by any F-AP, the cloud server directly transmits $F$ bits of their requested files to them.

In brief, the delivery scheme consists of three steps.
In the first step, the F-APs are divided into two groups. 
In the second step, the cloud server sends enough coded multicasting messages to the F-APs in the first group.
In the last step, the cloud server sends $F$ bits of the requested files to the F-APs in the second group.
The proposed coded delivery scheme is presented in Algorithm 2.

\begin{algorithm}[t]
\caption{F-AP Grouping Based Coded Delivery Scheme}
\label{delivery}
\begin{algorithmic}[1]		
\State Split the F-AP index set ${\cal K}$ into two subsets ${\cal K}_1$ and ${\cal K}_2$ based on $\boldsymbol{d}$;
\State $threshold \gets F(1-\frac{M}{N_0})$;
\State $sum \gets 0$;
\For{$j=|{\cal K}_1|,|{\cal K}_1|-1,\cdots,1$}
	\State $length \gets |W_{i,\mathcal{S}}^{'}|$, for all $\mathcal{S} \subseteq {\cal K}_1 : |\mathcal{S}| = j-1$;
	\State $increment \gets length \times \binom{|{\cal K}_1|-1}{j-1}$; 
	\State $sum_{new} \gets sum + increment$;
	\If{$sum + increment < threshold$}
		\State $ratio \gets length$;
	\Else
		\State $ratio \gets \frac{threshold - sum}{\binom{|{\cal K}_1|-1}{j-1}}$;
	\EndIf
	\For{${\cal P} \subseteq {\cal K}_1 : |{\cal P}| = j$}
		\State The cloud server transmits $ratio$ bits of the coded multicasting message $\oplus_{k \in {\cal P}} W^{'}_{d_k, {\cal P} \backslash \{ k\}}$ to the F-APs in $\cal P$;
	\EndFor
	\If{$sum_{new} \geq threshold$}
		\State break;
	\EndIf
	\State $sum \gets sum_{new}$;
\EndFor
\For{$k \in \mathcal{K}_2$}
	\State The cloud server transmits $F$ bits of $W_{d_k}^{'}$ to F-AP $k$ directly;
\EndFor
\end{algorithmic}
\end{algorithm}

\subsection{Performance Analysis}

In this subsection, we first derive a closed-form expression of the average fronthaul rate. Then, we optimize the parameters related to the placement scheme based on the derived closed-form expression. 

\subsubsection{Average Fronthaul Rate}
Let $p_0$ denote the probability that an F-AP requests a file from the cached group.
Then, it can be calculated as follows:
\begin{equation}
	p_0 = \sum_{j=1}^{N_0} p_j = \frac{\sum_{j=1}^{N_0} j^{- \alpha}}{\sum_{n=1}^N n^{- \alpha}}.
	\label{p0N0}
\end{equation}
According to the delivery scheme, ${\cal K}$ is split into ${\cal K}_1$ and ${\cal K}_2$ based on $\boldsymbol{d}$.
Let $k_1$ denote the number of elements in ${\cal K}_1$, i.e., $k_1=|{\cal K}_1|$.
It can be seen that $k_1$ is a binomial distributed random variable \cite{li2016traffic}. 
It is related to the popularity distribution and the file split point, and its probability mass function can be calculated as follows:
\begin{equation}
	\label{pmk}
	{\rm Pr} \left\{k_1=k\right\} = \binom{K}{k} p_0^k (1-p_0)^{K-k}, \quad k \in {\cal K}.
\end{equation}
When $k_1 = k$, it means that there are $k$ requests for files from the cached group and $K-k$ requests for files from the uncached group.

Let $p$ denote the probability that a particular bit of the coded file $W_i^{'}$ is cached in any single F-AP, where $i \in {\cal N}_1$. 
According to the placement scheme, $p$ can be calculated as
\begin{equation}
	p = \frac{MF/N_0}{F/r} = \frac{Mr}{N_0}.
\end{equation}
Let $[a:b]$ denote the set of integers $\{a,a+1,\ldots,b\}$ for $a \leq b$.
For a subset of F-APs ${\cal S} \in {\cal K}$, let $W^{'}_{i,s}(k)$ denote the set of bits of $W^{'}_i$ that are exclusively cached by F-APs in ${\cal S}$, where $s=|{\cal S}|$, ${\cal S} \subseteq [1:k]$ and $i \in {\cal N}_1$.
Then, the size of $W^{'}_{i,s}(k)$ can be calculated as
\begin{equation}
	|W^{'}_{i,s}(k)| =  \frac{F}{r} p^{s} (1-p)^{k-s}  = \frac{F}{r} \left(\frac{Mr}{N_0}\right)^s  \left(1-\frac{Mr}{N_0}\right)^{k-s}, k \in {\cal K}.
\end{equation}

Let $r_1$ denote the fronthaul rate required for meeting the $k$ requests for files from the cached group, and $r_2$ the fronthaul rate required for meeting the $K-k$ requests for files from the uncached group. 
When $k=0$, none of the F-APs request files from the cached group, so $r_1 = 0$. 
When $k>0$ and $M \geq N_0$, since each F-AP caches $M/N_0$ bits of the files from the cached group and $M/N_0 \geq 1$, each F-AP in ${\cal K}_1$ can reconstruct its requested file using the contents in its cache, so $r_1 = 0$.
When $k>0$ and $M < N_0$, the cloud server needs to transmit enough coded multicasting messages to F-APs in ${\cal K}_1$. 
If $|W^{'}_{i,k-1}(k)| < (1-M/N_0)F$, $r_1$ can be calculated as
\begin{equation}
	r_1 = \sum_{s=s_k+1}^{k-1} \binom{k}{s+1} \frac{|W^{'}_{i,s}(k)|}{F} + \eta_k \binom{k}{s_k+1} \frac{|W^{'}_{i,s_k}(k)|}{F},
\end{equation}
where $s_k$ and $\eta_k$ are the solutions of the following equation
\begin{equation}
	\sum_{s=s_k+1}^{k-1} \binom{k-1}{s} \frac{|W^{'}_{i,s}(k)|}{F} + \eta_k \binom{k-1}{s_k} \frac{|W^{'}_{i,s_k}(k)|}{F} = 1-\frac{M}{N_0},	
\end{equation}
with $s_k \in [0:k-1]$ and $\eta_k \in (0,1]$.
If $|W^{'}_{i,k-1}(k)| \geq (1-M/N_0)F$, $r_1$ can be calculated as
\begin{equation}
r_1 = 1-\frac{M}{N_0}.
\end{equation}
Therefore, we have:
\begin{equation}
\label{R1}
\begin{split}
r_1(k,M,N_0,r) 
= \left\{
\begin{array}{ll}
\sum_{s=s_k+1}^{k-1} \binom{k}{s+1} \mu_k(s+1) + \eta_k \binom{k}{s_k+1} \mu_k(s_k+1),\\  
\qquad \qquad \; \; M<N_0, k>0, \mu_k(k)<1-\frac{M}{N_0},\\ 
1-\frac{M}{N_0}, \quad \; \; M<N_0, k>0, \mu_k(k) \geq 1-\frac{M}{N_0},\\
0, \qquad \quad \: \; \; {\rm otherwise},
\end{array}
\right.
\end{split}
\end{equation}
where $\mu_m(x) =  \frac{1}{r} \left(\frac{Mr}{N_0}\right)^{x-1} \left(1-\frac{Mr}{N_0}\right)^{m-x+1}$.
According to the delivery scheme, the cloud server directly transmits $F$ bits of the requested file to F-APs in ${\cal K}_2$, so $r_2$ can be calculated as follows:
\begin{equation}
\label{R2}
	r_2 (K-k) = K-k. 
\end{equation}

In our proposed scheme, a placement strategy $\varphi$ is completely determined when the values of $N_0$ and $r$ are determined.
Therefore, the closed-form expression of the average fronthaul rate under placement strategy $\varphi$ can be calculated as follows:
\begin{equation}
\begin{split}
		\overline{R}_{\varphi} (K,M,N) &=\overline{R} (K,M,N,N_0,r) \\
		&= \sum_{\boldsymbol{d} \in \Theta} \left( \prod_{k=1}^K p_{d_k} \right) R_{\boldsymbol{d}}(K,M,N,N_0,r) \\
		&= \sum_{k=0}^{K} {\rm Pr} \left\{ k_1=k \right\} \left( r_1(k,M,N_0,r) + r_2(K-k) \right),
\end{split}
\end{equation}
where ${\rm Pr} \left\{ k_1=k \right\}$, $r_1(k,M,N_0,r)$ and $r_2(K-k)$ are given in (\ref{pmk}), (\ref{R1}) and (\ref{R2}), respectively.

\subsubsection{Parameters Optimization}
To optimize the parameters related to the placement scheme, problem (\ref{optimization1}) can be rewritten as follows:
\begin{subequations}
\begin{align}
	\min_{N_0,r} \; \: &\overline{R} (K,M,N,N_0,r) \tag{14} \\
	\textrm{s.t.} \quad &1 \leq N_0 \leq N, \quad N_0 \in {\cal Z}, \tag{14a} \\
	&0 < r \leq 1, \quad r \in {\cal R}, \tag{14b} 
\end{align}
\label{optimization2}%
\end{subequations} 
where $\cal Z$ is the set of all integers and $\cal R$ is the set of all real numbers.
 
Problem (\ref{optimization2}) is a mixed integer nonlinear programming problem. 
Due to the NP-hard property of this problem, it requires exponential computational complexity for traditional search methods to obtain the globally optimal solution.
In order to reduce the complexity of solving problem (\ref{optimization2}), we first discretize the value of the MDS code rate $r$. 
Let ${\cal A} = \{1/L,2/L,\ldots,1\}$, and we assume that $r \in {\cal A}$.
At the same time, as mentioned before, $r_1 = 0$ when $M \geq N_0$, so we have 
\begin{equation}
\begin{split}
	\overline{R} (K,M,N,N_0,r) &= \sum_{k=0}^{K} {\rm Pr} \left\{ k_1=k \right\} r_2(K-k) \\
   	&= K (1-p_0) \sum_{k=0}^{K-1} \binom{K-1}{k} p_0^k (1-p_0)^{K-1-k} \\
   	&= K (1-p_0), \quad N_0 \leq M.
\end{split} 
\end{equation}
According to (\ref{p0N0}), $p_0$ increases monotonically with $N_0$, so the average fronthaul rate decreases monotonically with $N_0$.
Therefore, for $N_0 \in [1:M]$, we have
\begin{equation}
	\overline{R} (K,M,N,M,r) \leq \overline{R} (K,M,N,N_0,r).
	\label{N0range}
\end{equation}
Based on (\ref{N0range}), the range of $N_0$ can be further limited to $[M:N]$.
Therefore, we transform problem (\ref{optimization2}) into the following optimization problem:
\begin{subequations}
\begin{align}
	\min_{N_0,r} \; \: &\overline{R} (K,M,N,N_0,r) \tag{17} \\
	\textrm{s.t.} \quad &M \leq N_0 \leq N, \quad N_0 \in {\cal Z}, \tag{17a} \label{rangeM} \\
	&0 < r \leq 1, \quad r \in {\cal A}. \tag{17b} \label{ranger}
\end{align}
\label{optimization3}%
\end{subequations}

Problem (\ref{optimization3}) is a pure integer programming problem, and we can traverse all possible value combinations of $r$ and $N_0$ under constraint (\ref{rangeM}) and (\ref{ranger}) to solve this problem and obtain the sub-optimal MDS code rate $r^{\star}$ and the sub-optimal file split point $N_0^{\star}$. 
The time complexity of this traversal algorithm is O$(L(N-M))$.

\section{Simulation Results}

In this section, the performance of our proposed scheme is evaluated via simulations. The parameters are set as follows: $F = 1$ Gb, $N=100$, $K=15$, $M=12$, $\alpha=0.8$, $L=100$. 

For performance comparison, we adopt four different caching schemes as baselines.
Baseline 1 refers to the traditional least-frequently-used (LFU) caching scheme, where the $M$ most popular files are entirely placed in each F-AP's cache.
Baseline 2 refers to the basic decentralized coded caching scheme in \cite{6807823}, where an $M/N$ portion of each file is placed in each F-AP's cache.
Baseline 3 refers to the multigroup decentralized coded caching scheme proposed in \cite{niesen2016coded}, where files are divided into multiple groups and files in the same group have similar popularity.
Baseline 4 refers to the random-least-frequently-used (RLFU) caching scheme proposed in 
\cite{ji2017order}, where files are divided into two groups but are not encoded by MDS
codes. 

In the following figures, we show the average fronthaul rate of different schemes.
For the baseline schemes that do not have a closed-form expression of the average fronthaul rate, we calculate the mean fronthaul rate during 2000 timeslots and use it to represent the average fronthaul rate. 
In each timeslot, a request vector is randomly generated according to the popularity distribution.
For baseline 3, we use the upper bound given in \cite{niesen2016coded}.

\begin{figure}[t]
	\centering
	\includegraphics[width=0.40\textwidth]{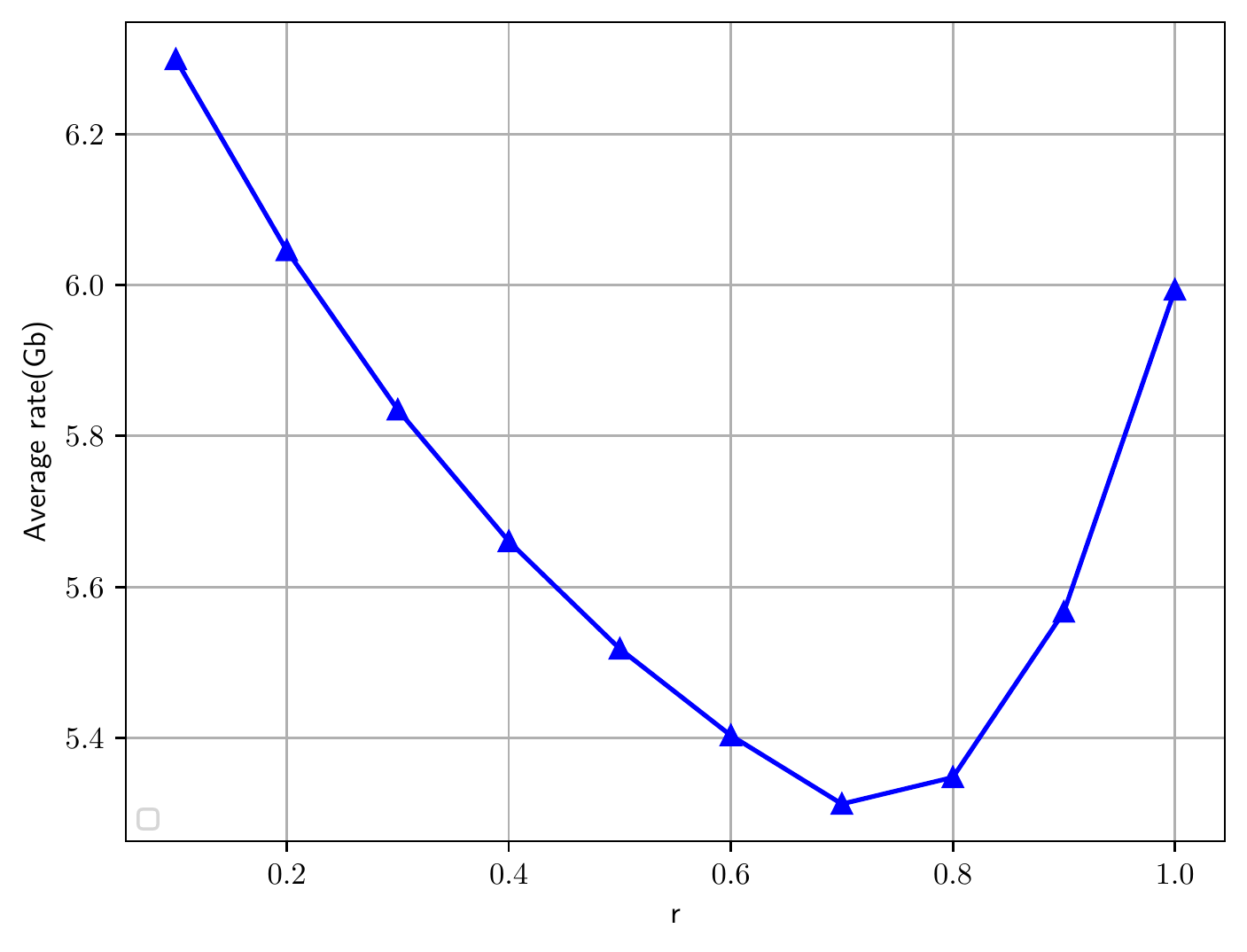}
	\caption{Average rate versus the MDS code rate $r$ with $K=15$, $M=12$.}
	\label{R_vs_r}
\end{figure}

Fig. \ref{R_vs_r} shows the effect of the MDS code rate, i.e., $r$, on the average fronthaul rate of the proposed scheme with $K=15$ and $M=12$. 
We consider 10 discrete values of $r$ ranging from 0.1 to 1. 
Note that when $r=1$, the size of the MDS coded file is equal to that of the original file, and we can assume that files are not encoded by MDS codes in this case. 
It can be seen that the optimal rate is 0.7, and the corresponding average fronthaul rate is approximately $12\%$ smaller than the one corresponding to $r=1$. 
These observations show that selecting an appropriate MDS code rate in the placement phase can effectively reduce the fronthaul rate. 

\begin{figure}[t]
	\centering
	\includegraphics[width=0.40\textwidth]{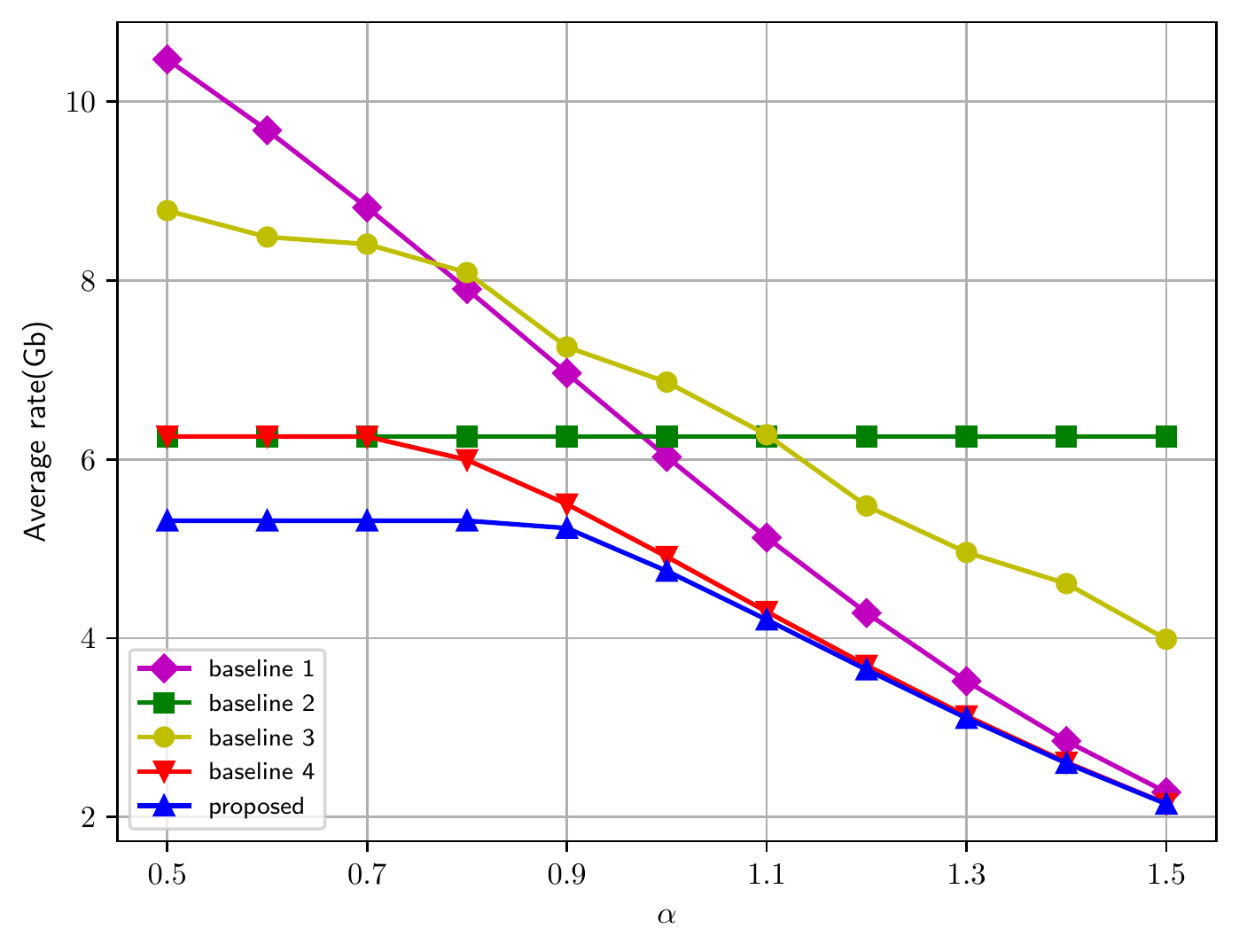}
	\caption{Average rate versus the Zipf distribution parameter $\alpha$ with $K=15$, $M=12$.}
	\label{R_vs_a}
\end{figure}

In Fig. \ref{R_vs_a}, we show the average fronthaul rate as a function of the Zipf distribution parameter $\alpha$ with $K=15$ and $M=12$.
It can be seen that as $\alpha$ increases, the average fronthaul rate of each scheme decreases except baseline 2. 
The reason is that baseline 2 treats files as having uniform popularity, while the proposed scheme and other baseline schemes preferably cache the most popular files according to the popularity distribution.
It can also be seen that as $\alpha$ increases, the performance gap between the proposed scheme and baseline 4 becomes smaller. 
The reason is that this performance gap is due to the fact that files are encoded by MDS codes in our proposed scheme, and the gain brought by MDS codes depends on the number of coded multicasting transmissions.
As $\alpha$ increases, each F-AP caches more bits of the most popular files, while the number of different files cached is reduced, so the number of coded multicasting transmissions is reduced, and the gain brought by MDS codes becomes smaller.

\begin{figure}[t]
	\centering
	\includegraphics[width=0.40\textwidth]{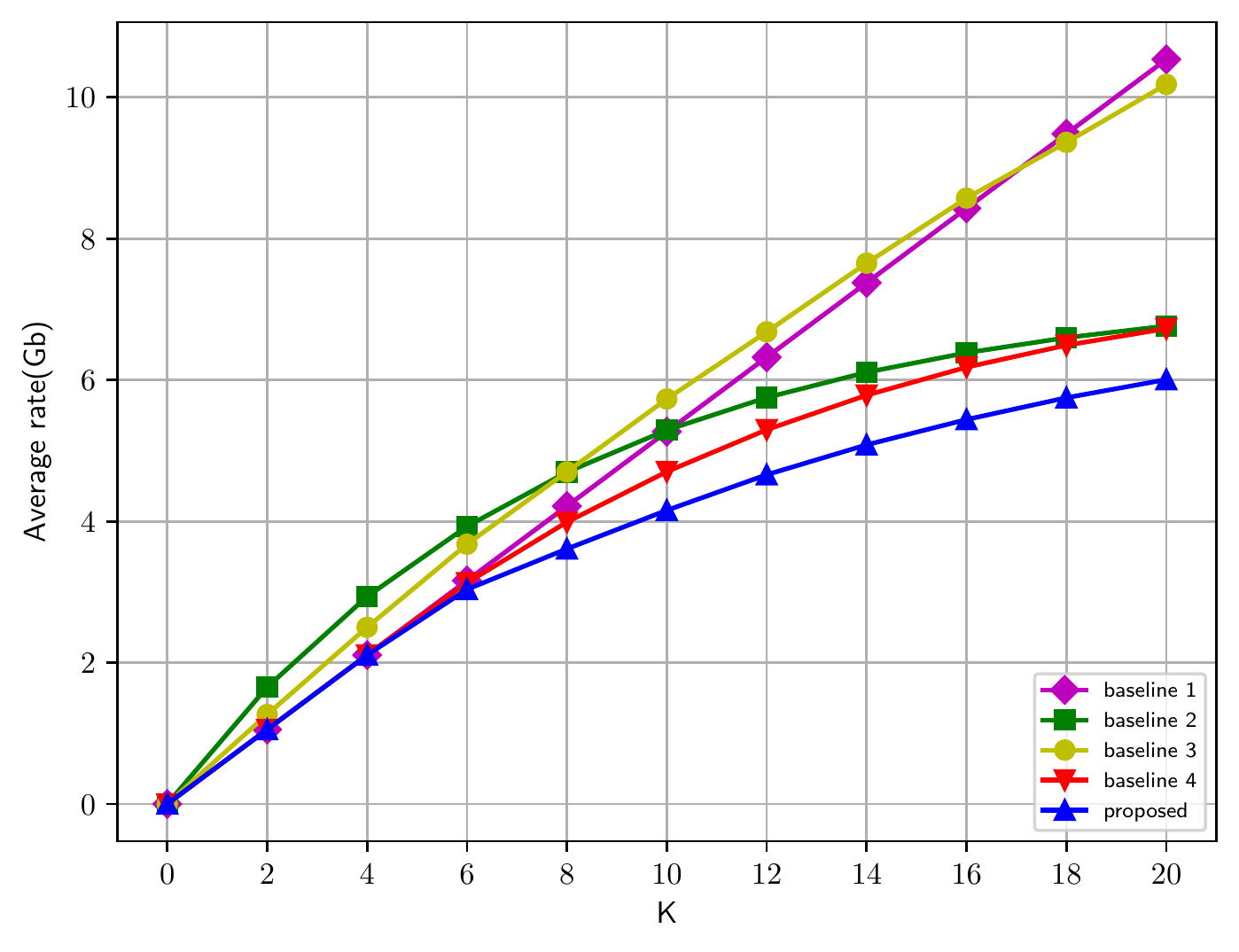}
	\caption{Average rate versus the number of F-APs $K$ with $M=12$.}
	\label{R_vs_K}
\end{figure}

In Fig. \ref{R_vs_K}, we show the effect of the number of F-APs, i.e., $K$, on the average fronthaul of each scheme with $M=12$.
As shown, our proposed scheme significantly reduces the fronthaul rate compared to the four baselines. 
With $K$ increasing, the cloud server needs to deal with more file requests, and the average fronthaul rates of all schemes increase. 
It can be observed that the performance gap between baseline 4 and our proposed scheme becomes larger as $K$ increases. 
The reason is that the cloud server does not need to send all the coded multicasting messages that are useful for subsets of F-APs in our proposed scheme due to the property of MDS codes. 
At the same time, the total size of the multicasting messages that do not need to be sent becomes larger as the number of file requests increases, so the gain brought by MDS codes becomes larger.

\begin{figure}[t]
	\centering
	\includegraphics[width=0.40\textwidth]{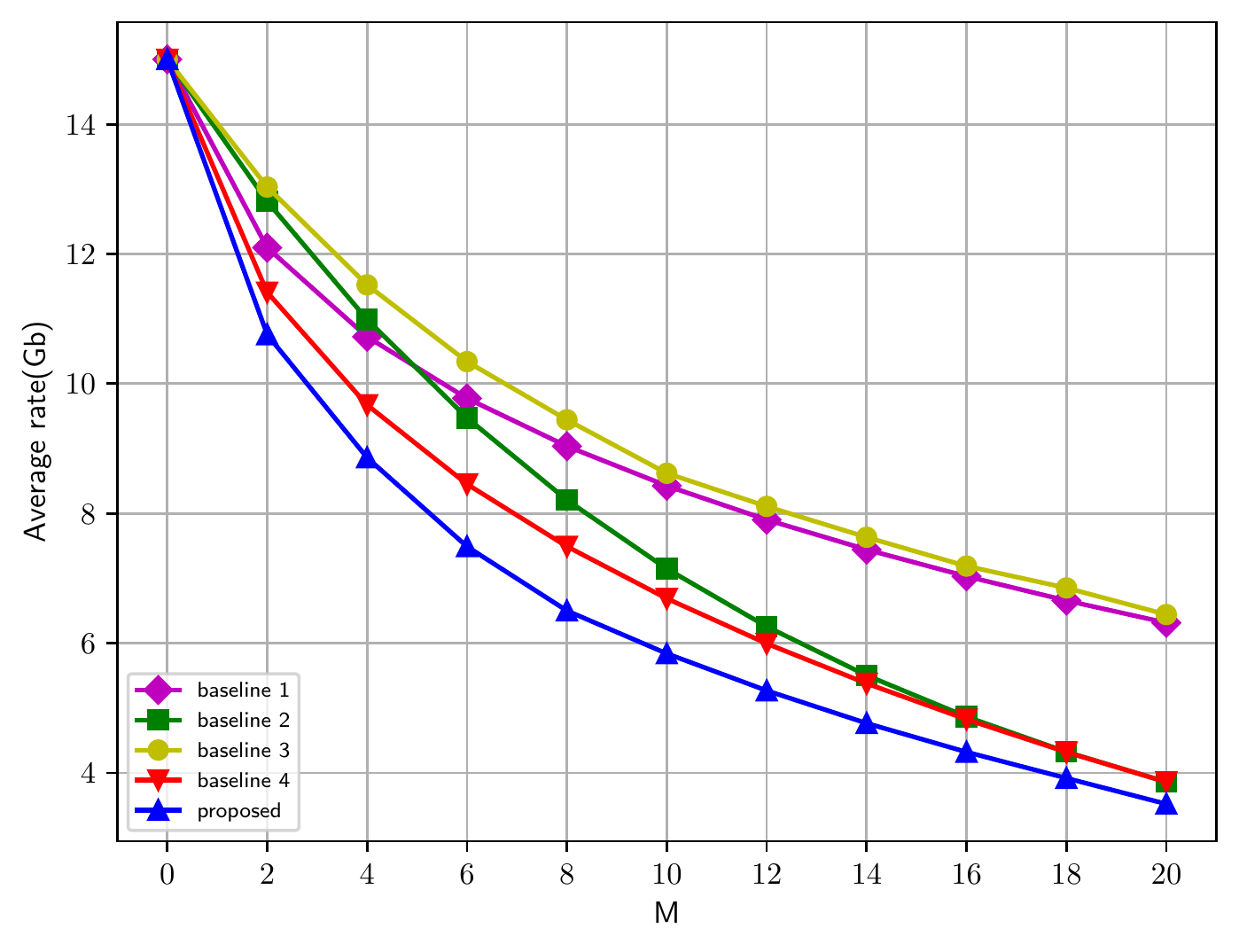}
	\caption{Average rate versus the cache capacity $M$ with $K=15$.}
	\label{R_vs_M}
\end{figure}

In Fig. \ref{R_vs_M}, we show the effect of the cache capacity of each F-AP, i.e., $M$, on the average fronthaul of each scheme with $K=15$.
With $M$ increasing, each F-AP caches more bits of files, and the average fronthaul rate of all schemes decreases. 
It can be observed that our proposed scheme further reduces the average fronthaul rate by approximately $20\%$ compared to baseline 2. 
This is due to the fact that we jointly optimize the file grouping strategy and the selection of the MDS code rate in our proposed scheme.
It can also be observed that our proposed scheme further reduces the average fronthaul rate by approximately $13\%$ compared to baseline 4. 
This is due to the fact that files are not encoded by MDS codes in baseline 4, while files are encoded by the MDS code whose rate is elaborately selected in our proposed scheme.

\section{Conclusions}

In this paper, we have proposed an MDS codes based group coded caching scheme in F-RANs under nonuniform file popularity. 
Specifically, we have designed an MDS codes and file grouping based coded placement scheme and an F-AP grouping based coded delivery scheme.
With the MDS code rate and the file split point elaborately selected, our proposed scheme can fully utilize the gains brought by the reconstruction properties of MDS codes and file grouping. 
Simulation results have shown that our proposed scheme can greatly reduce the fronthaul rate, especially when the number of F-APs is large. 
 
\section*{Acknowledgments}

This work was supported in part by the Natural Science Foundation of China under Grant 61971129, the Natural Science Foundation of Jiangsu Province under Grant BK20181264,
the Shenzhen Science and Technology Program under Grant KQTD20190929172545139 and JCYJ20180306171815699, and the National Major Research and Development Program of China under Grant 2020YFB1805005.

\bibliographystyle{IEEEtran}
\bibliography{paperREF}

\end{document}